\documentclass[12pt,a4paper,twoside]{article}
\usepackage[dvips]{graphics} 
\usepackage{cite}  
\usepackage{epsfig}
\usepackage{calc,ifthen}
\newcommand{\qqbar}  {\ensuremath{\mathrm{q\overline{q}}}}
\newcommand{\epem}   {\ensuremath{\mathrm{e^+e^-}}}

\newcommand{\zzero}  {\ensuremath{\mathrm{Z^0}}}

\newcommand{\evis}   {\ensuremath{E_{\mathrm{vis}}}}

\newcommand{\ycut}   {\ensuremath{y_{\mathrm{cut}}}}

\newcommand{\ppbar}   {\ensuremath{\mathrm{p\overline{p}}}}
\newcommand{\degs}   {\ensuremath{^\circ}}
\newcommand{\lam}   {\ensuremath{\Lambda}}
\newcommand{\kzero}   {\ensuremath{\mathrm{K^0_S}}}
\newcommand{\fthet}   {\ensuremath{f_{\Theta}}}
\newcommand{\fcut}   {\ensuremath{f^{\mathrm{cut}}_{\Theta}}}
\newcommand{\pur}   {\ensuremath{\rho_\mathrm{g}}}
\newcommand{\pnmix}   {\ensuremath{\rho_\mathrm{n.mix}}}

\newcommand{\rtag}   {\ensuremath{{\cal{P}}_\mathrm{tag}}}
\newcommand{\rgqi}   {\ensuremath{{\cal{R}}^i_{\mathrm{gq}}}}
\newcommand{\rgqch}   
{\ensuremath{{\cal{R}}^\mathrm{ch}_{\mathrm{gq}}}}
\newcommand{\rgql}   
{\ensuremath{{\cal{R}}^{\Lambda}_{\mathrm{gq}}}}
\newcommand{\rgqk}   
{\ensuremath{{\cal{R}}^{\mathrm{K^0_s}}_{\mathrm{gq}}}}
\newcommand {\gevc}      {\ensuremath{\mathrm{GeV}/c}}
\newcommand {\gevcc}      {\ensuremath{\mathrm{GeV}/c^2}}

\newcommand{\pkzero}   {\ensuremath{p_{\mathrm{K^0_s}}}}
\newcommand{\plam}   {\ensuremath{p_{\Lambda}}}
\newcommand{\eps}    {\ensuremath{\varepsilon}}

\newcommand{\roots} {\ensuremath{\sqrt{s}}} 

\newcommand{\refnum} {CERN-EP/98-058}
\newcommand{\Date} {30th March 1998}

\newcounter{hours}\newcounter{minutes}

\newcommand{\Printtime}{%
  \setcounter{hours}{\time/60}%
  \setcounter{minutes}{\time-\value{hours}*60}%
  \ifthenelse{\value{hours}<10}{0}{}\thehours:%
  \ifthenelse{\value{minutes}<10}{0}{}\theminutes}
\newcommand{\ywin}   {\ensuremath{y_{\mathrm {win}}}}
\newcommand{\ks}     {\ensuremath{{\rm K}^0_{\rm S}}}
\newcommand{\tjet}   {{three-jet event}}
\newcommand{\durham} {\ensuremath{k_{\perp}}}
\newcommand{\jt}     {{\sc Jetset}}
\newcommand{\hw}     {{\sc Herwig}}

\begin{document}
%
\begin{titlepage}
%
\vspace*{-2cm}
\begin{center}
{\large EUROPEAN LABORATORY FOR PARTICLE PHYSICS}
\end{center}

\bigskip
\begin{flushright}
    \refnum\\
    \Date
\end{flushright}
\vskip 1.cm
 
\begin{center} \Large\bf\boldmath
Production of \ks\ and \lam\ in 
Quark  and Gluon Jets from \zzero\ Decay
\end{center}
 
\bigskip
\begin{center}
{\Large The OPAL Collaboration}
\end{center}
\bigskip\bigskip
  
\begin{abstract}
 
  The production of \ks\ mesons and \lam\ baryons in quark and gluon
  jets has been investigated using two complementary techniques. In
  the first approach, which provides high statistical accuracy, jets
  were selected using different jet finding algorithms and ordered
  according to their energy. Production rates were determined taking
  into account the dependences of quark and gluon compositions as a
  function of jet energy as predicted by Monte Carlo models.
  Selecting {\tjet}s with the \durham\ (Durham) jet finder (\ycut\ =
  0.005), the ratios of \ks\ and \lam\ production rates in gluon and
  quark jets relative to the mean charged particle multiplicity were
  found to be $1.10 \pm 0.02 \pm 0.02$ and $1.41 \pm 0.04 \pm 0.04$,
  respectively, where the first uncertainty is statistical and the
  second is systematic.  In the second approach, a new method of
  identifying quark jets based on the collimation of energy flow
  around the jet axis is introduced and was used to anti-tag gluon
  jets in symmetric (Y-shaped) three-jet events.
  Using the cone jet finding algorithm with a cone
  size of 30\degs, the ratios of relative production rates in gluon
  and quark jets were determined to be $0.94 \pm 0.07 \pm 0.07$ for
  \kzero\ and $1.18 \pm 0.10 \pm 0.17$ for \lam.  The results of both
  analyses are compared to the predictions of Monte Carlo models.
 
\end{abstract}

\bigskip 

\bigskip \bigskip \bigskip
\begin{center}
{\large Submitted to European Physics Journal C.}
\end{center}


\end{titlepage}

\newpage

\begin{center}{\Large        The OPAL Collaboration
}\end{center}\bigskip
\begin{center}{
K.\thinspace Ackerstaff$^{  8}$,
G.\thinspace Alexander$^{ 23}$,
J.\thinspace Allison$^{ 16}$,
N.\thinspace Altekamp$^{  5}$,
K.J.\thinspace Anderson$^{  9}$,
S.\thinspace Anderson$^{ 12}$,
S.\thinspace Arcelli$^{  2}$,
S.\thinspace Asai$^{ 24}$,
S.F.\thinspace Ashby$^{  1}$,
D.\thinspace Axen$^{ 29}$
% ...
}\end{center}\bigskip
\bigskip
$^{  1}$School of Physics and Astronomy, University of Birmingham,
Birmingham B15 2TT, UK
\newline
$^{  2}$Dipartimento di Fisica dell' Universit\`a di Bologna and INFN,
I-40126 Bologna, Italy
\newline
$^{  3}$Physikalisches Institut, Universit\"at Bonn,
D-53115 Bonn, Germany
\newline
$^{  4}$Department of Physics, University of California,
Riverside CA 92521, USA
\newline
$^{  5}$Cavendish Laboratory, Cambridge CB3 0HE, UK
\newline
$^{  6}$Ottawa-Carleton Institute for Physics,
Department of Physics, Carleton University,
Ottawa, Ontario K1S 5B6, Canada
\newline
$^{  7}$Centre for Research in Particle Physics,
Carleton University, Ottawa, Ontario K1S 5B6, Canada
\newline
$^{  8}$CERN, European Organisation for Particle Physics,
CH-1211 Geneva 23, Switzerland
\newline
$^{  9}$Enrico Fermi Institute and Department of Physics,
University of Chicago, Chicago IL 60637, USA
\newline
$^{ 10}$Fakult\"at f\"ur Physik, Albert Ludwigs Universit\"at,
D-79104 Freiburg, Germany
\newline
$^{ 11}$Physikalisches Institut, Universit\"at
Heidelberg, D-69120 Heidelberg, Germany
\newline
$^{ 12}$Indiana University, Department of Physics,
Swain Hall West 117, Bloomington IN 47405, USA
\newline
$^{ 13}$Queen Mary and Westfield College, University of London,
London E1 4NS, UK
\newline
$^{ 14}$Technische Hochschule Aachen, III Physikalisches Institut,
Sommerfeldstrasse 26-28, D-52056 Aachen, Germany
\newline
$^{ 15}$University College London, London WC1E 6BT, UK
\newline
$^{ 16}$Department of Physics, Schuster Laboratory, The University,
Manchester M13 9PL, UK
\newline
$^{ 17}$Department of Physics, University of Maryland,
College Park, MD 20742, USA
\newline
$^{ 18}$Laboratoire de Physique Nucl\'eaire, Universit\'e de Montr\'eal,
Montr\'eal, Quebec H3C 3J7, Canada
\newline
$^{ 19}$University of Oregon, Department of Physics, Eugene
OR 97403, USA
\newline
$^{ 20}$Rutherford Appleton Laboratory, Chilton,
Didcot, Oxfordshire OX11 0QX, UK
\newline
$^{ 22}$Department of Physics, Technion-Israel Institute of
Technology, Haifa 32000, Israel
\newline
$^{ 23}$Department of Physics and Astronomy, Tel Aviv University,
Tel Aviv 69978, Israel
\newline
$^{ 24}$International Centre for Elementary Particle Physics and
Department of Physics, University of Tokyo, Tokyo 113, and
Kobe University, Kobe 657, Japan
\newline
$^{ 25}$Institute of Physical and Environmental Sciences,
Brunel University, Uxbridge, Middlesex UB8 3PH, UK
\newline
$^{ 26}$Particle Physics Department, Weizmann Institute of Science,
Rehovot 76100, Israel
\newline
$^{ 27}$Universit\"at Hamburg/DESY, II Institut f\"ur Experimental
Physik, Notkestrasse 85, D-22607 Hamburg, Germany
\newline
$^{ 28}$University of Victoria, Department of Physics, P O Box 3055,
Victoria BC V8W 3P6, Canada
\newline
$^{ 29}$University of British Columbia, Department of Physics,
Vancouver BC V6T 1Z1, Canada
\newline
$^{ 30}$University of Alberta,  Department of Physics,
Edmonton AB T6G 2J1, Canada
\newline
$^{ 31}$Duke University, Dept of Physics,
Durham, NC 27708-0305, USA
\newline
$^{ 32}$Research Institute for Particle and Nuclear Physics,
H-1525 Budapest, P O  Box 49, Hungary
\newline
$^{ 33}$Institute of Nuclear Research,
H-4001 Debrecen, P O  Box 51, Hungary
\newline
$^{ 34}$Ludwigs-Maximilians-Universit\"at M\"unchen,
Sektion Physik, Am Coulombwall 1, D-85748 Garching, Germany
\newline
\bigskip\newline
$^{  a}$ and at TRIUMF, Vancouver, Canada V6T 2A3
\newline
$^{  b}$ and Royal Society University Research Fellow
\newline
$^{  c}$ and Institute of Nuclear Research, Debrecen, Hungary
\newline
$^{  d}$ and Department of Experimental Physics, Lajos Kossuth
University, Debrecen, Hungary
\newline
$^{  e}$ on leave of absence from the University of Freiburg
\newline

\newpage

\section{Introduction}
\label{intro}

In recent years, clear differences between quark and gluon jets have
been established experimentally
\cite{PR038,PR070,PR136,PR141,bib_qgdiffh,LEP-qg}.
These studies have mostly exploited the large samples of \zzero\ decay
events recorded at the CERN LEP collider, and have examined three-jet
events with a symmetric event topology favouring the direct comparison
of quark and gluon jets of the same energy produced in the same event
topology. In particular, gluon jets have been measured to have a
larger mean particle multiplicity, a softer fragmentation function and
a larger angular width than quark jets of the same energy. Comparisons
between jets produced in \epem\ and \ppbar\ collisions also indicate
that gluon jets are broader than quark jets \cite{PR097}. This is in
qualitative agreement with the predictions of perturbative QCD
\cite{bib_qcd}.  Perturbative QCD makes no explicit predictions for
the individual hadron species produced in these jets, but some QCD
Monte Carlo models of hadronization predict that the relative hadron
production rates in quark and gluon jets differ for different hadron
species.

Few experimental results are available on the production of identified
particles in quark and gluon jets.  Results from \epem\ annihilations
at center-of-mass (c.m.) energies around the $\Upsilon(1S)$ resonance
($\approx 10$~GeV) indicate that baryons are produced about 2.5 times
more copiously in direct $\Upsilon(1S)$ decays ($\Upsilon(1S)
\rightarrow$ ggg $\rightarrow$ hadrons) than in continuum events
($\epem \rightarrow \qqbar \rightarrow$ hadrons) whilst no such
enhancement was observed for mesons \cite{bib_argus_cleo}.  An OPAL
investigation~\cite{PN246} studied identified particle production in
jet topologies enriched in gluon and quark jets and found no
jet-dependent differences in the production of mesons and charged
particles (other than protons).  In contrast, baryons were found to be
produced more copiously in gluon-enriched jet samples.  The DELPHI
collaboration has reported measurements of K$^\pm$, K$^0$, p and
$\Lambda$ in secondary-vertex tagged quark and gluon jets in symmetric
(Y-shaped) events \cite{bib_qqgiddel}. Within relatively large
statistical and systematic uncertainties, they find the ratio of
identified particle rates in quark and gluon jets to be consistent
with that for charged particles.  Recently, L3 has concluded that
production of K$^0$ and $\Lambda$ in both quark and gluon jets is well
modelled by string fragmentation~\cite{bib_qqgidl3}.

In this paper an experimental comparison of K$^0$ and $\Lambda$
production in quark and gluon jets is presented.  Two complementary
analyses of data recorded with the OPAL detector at LEP are presented,
using different approaches to identify quark and gluon jets.  One
analysis separates quark and gluon jets according to their energies,
giving a large sample of events, whilst the other selects a smaller
sample of tagged quark and gluon jets in symmetric (Y-shaped) events,
which allows for a simpler interpretation, but with larger
uncertainties.

In the first analysis (the `energy-based analysis') the production of
\ks\ mesons and \lam\ baryons\footnote{For simplicity \lam\ refers to
  both \lam\ and ${\overline{\Lambda}}$.} in quark and gluon jets was
investigated in three-jet events of different topologies selected with
the \durham\ (Durham)~\cite{bib_durham} or a cone~\cite{PR097} jet
finder.  The jets were ordered by their energy since the lowest energy
jets are mainly induced by gluons, and higher energy jets mainly by
quarks. Motivated by Monte Carlo investigations a similar energy
dependence for the production rates of all hadron species was assumed,
and the production rates of \kzero\ and \lam\ relative to those of
charged particles were determined.  The experimental relative rates
were corrected for the underlying mixture of quark and gluon jets,
allowing the \ks\ and \lam\ relative production rates in pure quark
and gluon jets to be obtained.

In the second analysis (the `Y-event analysis'), a comparison was made
of the absolute production rates of \kzero\ mesons and \lam\ baryons
in quark and gluon jets produced under the same conditions, embedded
in similar event topologies. Symmetric three-jet events were analysed,
where the two lower energy jets (assumed to be initiated by a quark
and a gluon) were produced at about 150\degs\ with respect to the
higher energy jet.  A sample of anti-tagged gluon jets containing
about 30\% of the symmetric event sample was isolated by means of a
new method of identifying the quark jets based on the observation that
light quark jets are more collimated than gluon jets \cite{PR136}. The
inclusive yield of charged particles in these jets was also measured,
allowing relative rates to be evaluated.  The production rates of the
particles in the lower energy jets of the inclusive symmetric sample
were determined, and a correction applied in order to obtain
measurements corresponding to pure samples of quark and gluon jets.

Section~\ref{detector} gives a description of the main features of the
OPAL detector, of the data and simulated event samples and of the
reconstruction algorithms for \kzero\ and \lam.
Section~\ref{eoanalysis} describes the details of the energy-based
analysis, and section~\ref{Yanalysis} the Y-event analysis.  The
results of both analyses are presented and discussed in section
\ref{discuss}, before the summary is given in section~\ref{summary}.

\section{The OPAL Detector and Data Samples}
\label{detector}

\subsection{The OPAL Detector}
\label{OPAL}

The OPAL detector is described in detail elsewhere \cite{PR021}. Of
most relevance for the present analyses are the tracking system and
the electromagnetic calorimeter. The tracking system consists of a
silicon microvertex detector, an inner vertex chamber, a large-volume
jet chamber and specialised chambers at the outer radius of the jet
chamber which improve the measurements in the $z$
direction\footnote{The coordinate system is defined so that $z$~is the
  coordinate parallel to the e$^-$ beam axis, $r$~is the coordinate
  normal to the beam axis, $\phi$~is the azimuthal angle around the
  beam axis and $\theta$~is the polar angle \mbox{with respect
    to~$z$.}} ($z$-chambers). The tracking system covers the region
$|\cos\theta|<0.95$ and is enclosed by a solenoidal magnet coil with
an axial field of~0.435~T.  The tracking detectors provide momentum
measurements of charged particles, and particle identification from
measurements of the ionisation energy loss, d$E$/d$x$.
Electromagnetic energy is measured by a lead-glass calorimeter located
outside the magnet coil, which covers $|\cos\theta|<0.98$.

\subsection{Data Samples}
\label{data}
The energy-based analysis, which is not statistics limited, used 2.8
million events recorded between 1992 and 1994 whilst the Y-event
analysis used the full OPAL data sample of about 4.2 million hadronic
events collected around the \zzero\ peak from 1990 to 1995.  The
procedures for identifying hadronic events using measurements of
tracks and electromagnetic energy are described
in~\cite{PR035}.  The criteria applied to select tracks and deposits
of electromagnetic energy (clusters) for the analyses were identical
to those in~\cite{PR136}.
Each accepted track and unassociated electromagnetic cluster was
considered to be a particle.  Tracks were assigned the pion mass and
electromagnetic clusters were assigned zero mass since they originate
mostly from photons.

To reduce background from non-hadronic decays of the \zzero\ and to
eliminate events in which a significant number of particles were lost
near the beam direction, both analyses required that $|\cos
(\theta_{\mathrm {thrust}})|<0.9$ ($\theta_{\mathrm {thrust}}$ is the
polar angle of the thrust axis); also, there had to be at least five
accepted tracks.  The residual background from all sources was
estimated to be less than~1\%.  In addition, for the energy-based
analysis, events were rejected if they contained tracks with measured
momentum greater than 60~\gevc, if the absolute value of the vector
sum of all selected particles $|\vec{p}_{\rm tot}|$ exceeded 30~\gevc,
or if the visible energy (the sum of the energies of all the accepted
tracks and clusters) was less than 40\% of the center-of-mass energy.

\subsection{Reconstruction of \ks\ and \lam}
\label{lamk0s}

The neutral strange \ks\ mesons and \lam\ baryons were reconstructed
by their decay channels \ks~$\rightarrow \pi^+ \pi^-$ and
\lam~$\rightarrow \pi^-$p.  The reconstruction algorithms, signal
definitions, efficiency corrections and background subtractions are
all described in \cite{PR126} and \cite{PR061}. Briefly, tracks of
opposite charge were paired and regarded as a secondary vertex
candidate if at least one track pair intersection in the plane
perpendicular to the beam axis satisfied the criteria of a neutral
two-body decay with the appropriate lifetime.  Each track pair passing
these requirements was refitted with the constraint that the tracks
originated from a common vertex, and background from photon
conversions was suppressed. For \lam\ candidates, information from
d$E$/d$x$ measurements was used to help identify the $\pi$ and p for
further background suppression.  Two sets of cuts are described in
\cite{PR061} for \lam\ identification.\footnote{The two sets of cuts
  are described as `method 1' and `method 2' in \cite{PR061}.}  For
the energy-based analysis, \lam\ candidates were reconstructed using
method~1, which is optimized to have good mass and momentum
resolution, while in the Y-event analysis the more efficient method~2
was employed to maximize the number of \lam\ candidates.

Candidates for \ks\ with momentum greater than 0.150~\gevc\ and with an
invariant mass in the range $0.3~\gevcc\ < m_{\pi\pi}< 0.8~\gevcc$ and
\lam\ candidates with momentum greater than 0.520~\gevc\ and with an
invariant mass in the range $1.08~\gevcc\ m_{\pi p}< 1.2~\gevcc$ were
retained for further analysis.  The \ks\ rates were determined by
fitting the mass spectrum with a third-order polynomial excluding a
signal region of $\pm 0.05~\gevcc$ around the nominal mass. The shape
of the background under the \lam\ signal was fitted using a function
of the form $(1-e^{-a({\rm m}_{\pi p}-1.077)}) \times(b\cdot {\rm
  m}_{\pi p}+c)$, excluding a signal window of $\pm 0.012~\gevcc$
around the nominal \lam\ mass.  For each particle species the entries
in the signal region were summed and the background as determined by
the fit was subtracted.  The efficiency of the identification
algorithms was determined as a function of candidate momentum from
Monte Carlo events, and subsequently used to correct the number of
observed signal events.

For the energy-based analysis, candidates for \lam\ and \kzero\ 
were formed and then supplied to the jet finder, whereas this
order was reversed in the case of the Y-events. Monte Carlo studies
showed that for the Y-event analysis there is no systematic effect 
due to the order of jet finding and particle reconstruction.

\subsection{Monte Carlo Event Samples}
\label{MC}

Samples of Monte Carlo hadronic events with a full simulation of the
OPAL detector \cite{gopal} and including initial state photon
radiation were used to evaluate the detector acceptance and
resolution, and to study the efficiency and purity of the quark and
gluon jet identification and the particle reconstruction algorithms.
In total, 7 million simulated events were available, of which 4
million were generated by \jt~7.4~\cite{bib_jt74} with fragmentation
parameters described in~\cite{PR141}, and 3 million generated by
\jt~7.3 with fragmentation parameters described in~\cite{PR070}.  The
\jt~7.4 events included updated particle decay tables and heavy meson
resonances and were processed using a more recent version of the
detector simulation compared to the \jt~7.3 sample. It should be noted
that there are significant differences in the simulation of baryon
production between the two \jt\ samples.

For comparison with the experimental results, the Monte Carlo models
\jt~7.4 and \hw~5.9\footnote{The fragmentation parameters of \hw~5.9
  were identical to those used by the OPAL tuned version of \hw~5.8
  \cite{PR158} with the exception of the parameter {\tt CLMAX} which
  was set to 3.75 in order to improve the description of the mean
  charged particle multiplicity value in inclusive hadronic \zzero\ 
  decays.}  \cite{bib_hw} were used.  The models both give a good
description of global event shapes and many inclusive particle
production rates, but differ in their description of the perturbative
phase and their implementation of the hadronization mechanism.

Tracks and clusters were selected in the Monte Carlo events, which
include detector simulation, in the same way as for the data: the
resulting four-vectors of particles are referred to as being at the
`detector level'.  Alternatively, Monte Carlo samples without
initial-state photon radiation or detector simulation were used, with
all charged and neutral particles with mean lifetimes greater than
$3\cdot10^{-10}$~s treated as stable.  The four-vectors of the
resulting particles are referred to as being at `generator level'. The
remaining quarks and gluons after the termination of the parton shower
in these events are referred to as being at the `parton level'.

\section{The Energy-Based Analysis}
\label{eoanalysis}

\subsection{Selection of Three-Jet Events}
\label{tjets}

Two different types of algorithms are commonly used for jet
definition: recombination jet finders and cone jet finders.  The
treatment of high momentum particles which lie close to the jet axes
is similar for both types of jet finder, but there are substantial
differences in the assignment to the jets of soft particles far from
the jet axes.  Recombination algorithms combine all particles into
jets, whereas cone finders do not associate particles outside the
cones.  The recombination jet finders iteratively combine pairs of
particles until the scaled resolution parameter ($y$) of all
subsequent pairings exceeds the jet resolution parameter \ycut. The
cone jet finder associates particles into jets that lie within a cone
of fixed half-angle $R$.  The cone axis is determined from the vector
sum of the momenta of the particles contained therein.

To illustrate the sensitivity of the analysis to event topology and
jet reconstruction algorithm, we selected three samples of three-jet
events with different average topologies, using either the \durham\ 
jet finding algorithm~\cite{bib_durham}, or the cone
algorithm~\cite{PR097}.  The main selection, containing the largest number
of three-jet events, was obtained using the \durham\ algorithm with a
resolution parameter, \ycut\ = 0.005. The second sample of three-jet
events (\ywin\ sample) was selected
from a window of $y$-values between 0.008 and 0.016, i.e.,
the event should be classified as three-jet for
some value of \ycut\ within this window.
  Monte Carlo
studies at the generator and parton level showed that hadronization
effects on the angles of jets selected in this sample are small. The
average resolution parameter \ycut\ for these events to be clustered
into three jets was about 0.01.  Finally, a third sample of {\tjet}s
was selected using the cone jet finder with the parameters $R$ =
0.7~rad and $\varepsilon$ = 7~GeV, where $\varepsilon$ is the minimum
energy contained in the jet cone.  These parameters were chosen to
give a good correspondence between jets reconstructed at the generator
and parton levels in Monte Carlo events \cite{PR097}.

Both jet finding algorithms were applied using reconstructed \ks\ and \lam\ 
candidates, accepted charged tracks (excluding the decay products of \ks\ and 
\lam) and the electromagnetic clusters not associated to tracks. 
Additional quality cuts were then applied to the selected three-jet events. 
Each jet was required to contain at least two charged tracks, to have
more than 5 GeV of visible energy and to lie in the polar region
$|\cos \theta|<$~0.9.  The sum of the angles between all three jets
had to exceed $358^{\circ}$ to eliminate non-planar events, and the
angle between the two lower energy jets was required to be larger than
$30^{\circ}$.  Finally, the jets in each event were each assigned a
calculated energy based on the measured jet directions and assuming
massless kinematics.

\begin{figure}[htb]
\begin{center}
\resizebox{\textwidth}{!}{\includegraphics{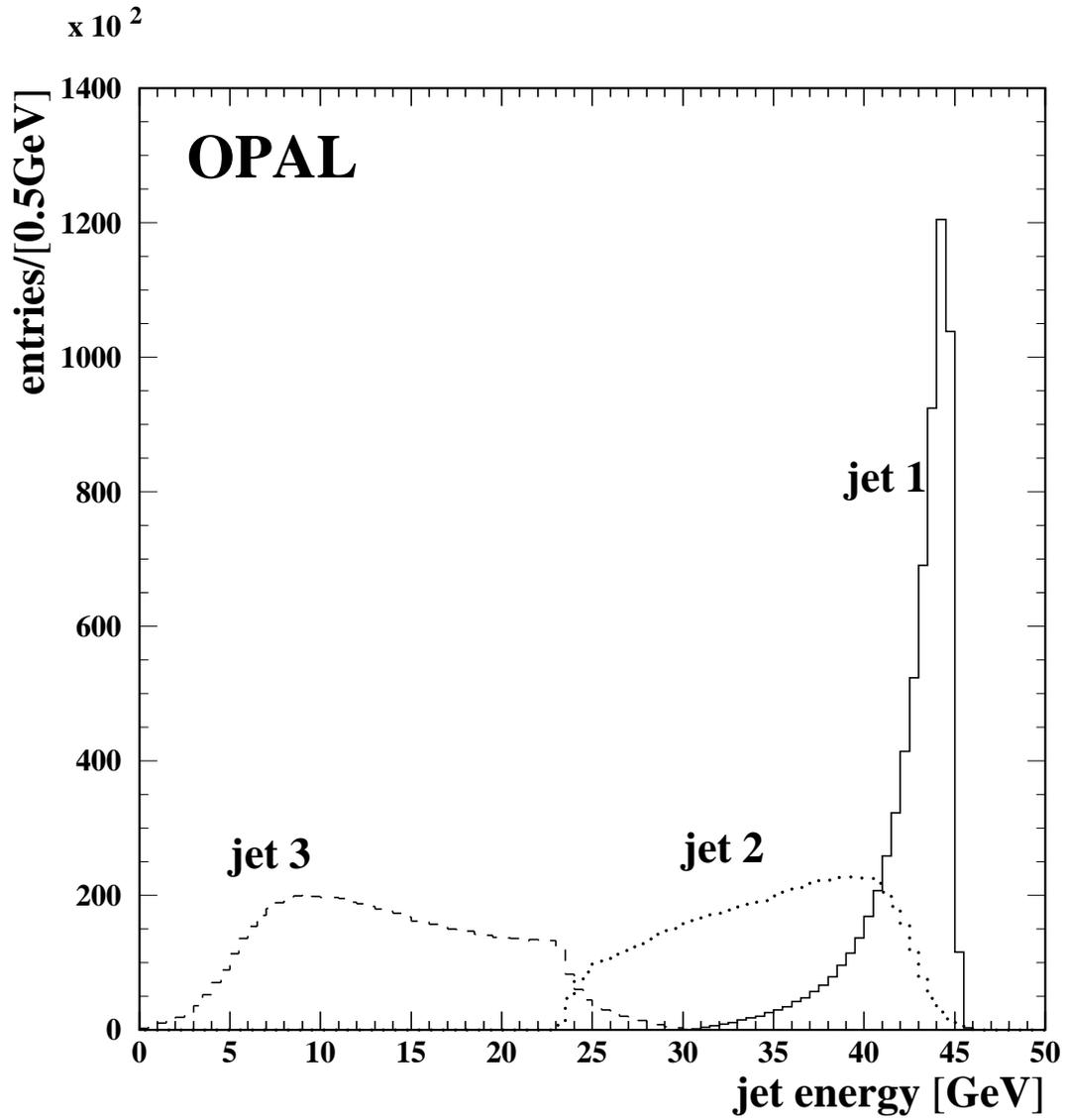}}
\end{center}
\caption
  {Jet energy distributions of three-jet events selected with 
   the \durham\ jet finder with the \ycut\ event selection.The jets
   are ordered according to their assigned energies.}
\label{fig_b1}
\end{figure}

After all cuts, $24.0\%$ of events were selected by the \ycut\ 
selection, $18.1\%$ by the \ywin\ selection, and $15.5\%$ by the 
cone selection.
The tighter selection criteria for the \ywin\ and the cone samples,
yielding a lower number of {\tjet}s, were partly compensated by an
improved reconstruction quality e.g., a better agreement of the jets
at the parton and generator level. In total, samples of about 500,000
{\tjet}s were retained for further analysis.  In figure~\ref{fig_b1}
the energy spectra of the jets in the \ycut\ sample is shown.
The jet energy distributions are more peaked in the \ycut\ selection
compared to the \ywin\ and cone selections. The relatively small
number of events containing a jet with energy below 7.5~GeV ($\approx
15\%$) were excluded from further analysis.

In general, the angular separation of the jets (jet topology), is
dependent on the event selection. In particular, the \ycut\ selection
gave the most collimated sample of events, with an average angle
between the two lowest energy jets of about 62\degs.  This can be
compared to average angles of about 70\degs\ for the \ywin\ sample and
about 76\degs\ for the cone selection.

\subsection{Determination of Jet Purities}
\label{purities}

\begin{figure}[htb]
\begin{center}
\resizebox{\textwidth}{!}{\includegraphics{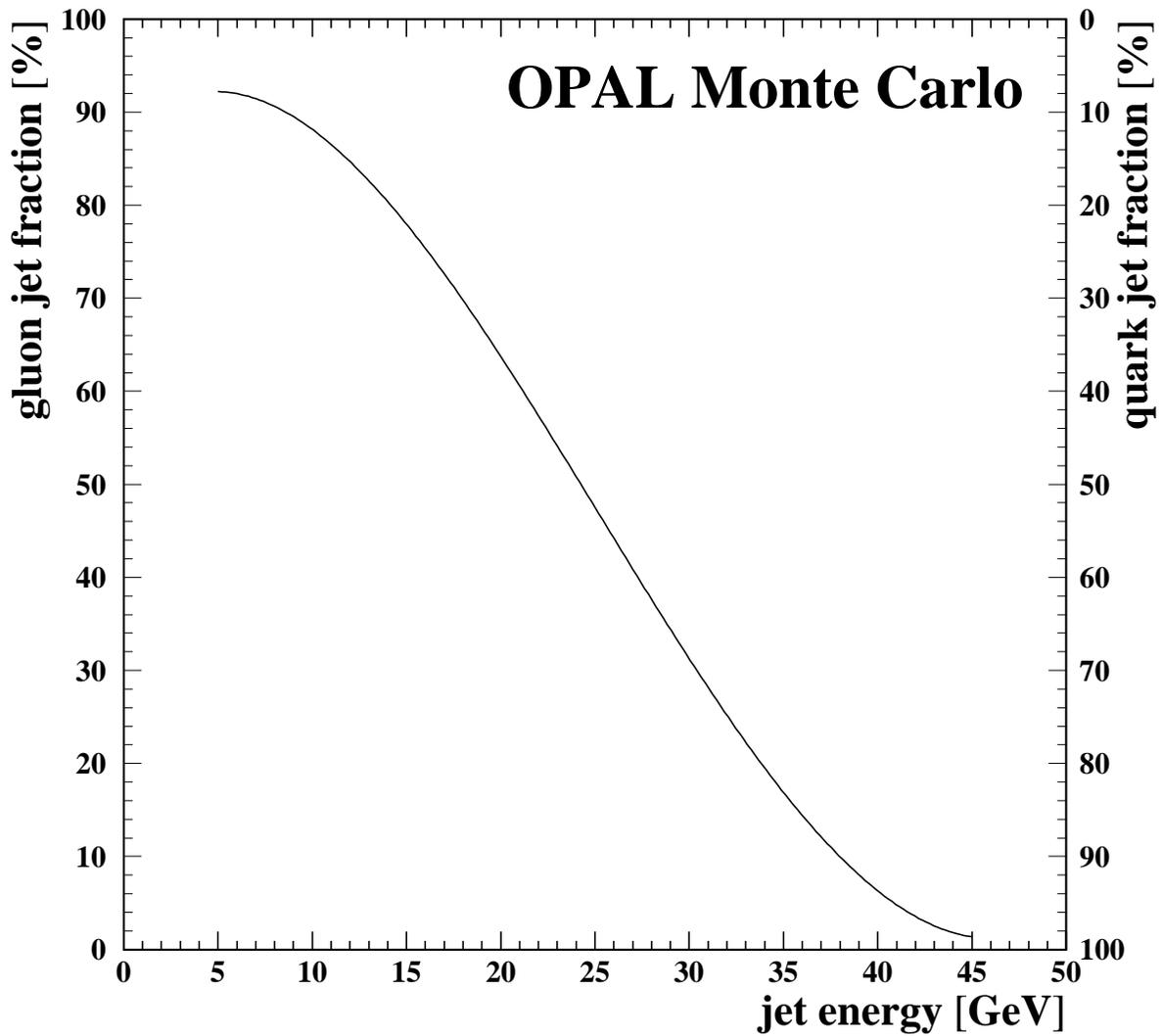}}
\end{center}
\caption
  {Purities of the reconstructed jets in Monte Carlo events 
   for the \ycut\ event sample.}
\label{fig_b2}
\end{figure}

Since the jets selected in any given energy interval were a mixture of
quark and gluon jets, \jt\ Monte Carlo events were used to determine
the quark and gluon jet content of the samples in the following
manner. The Monte Carlo events were selected in the same way as the
data events and were accepted if classified as three-jet events at
detector level. From the partons of these events exactly three jets
were reconstructed. The detector level jets closest in angle to the
parton level jets containing the primary quark and anti-quark were
considered to be the quark jets and the remaining jet the gluon jet.
The term quark (gluon) `jet purity' is defined to be the fraction of
jets in a given energy bin, initiated by quarks (gluons).  The
purities of the jets as a function of jet energy are shown in
figure~\ref{fig_b2}.  The lowest energy jet samples contained in
excess of 85\% gluons, whereas the high energy samples were composed
of over 95\% quark jets.  The purity distributions as a function of
jet energy determined with the \durham\ algorithm in the \ycut\ and
\ywin\ samples agree to within 1\%. The gluon jet purity in the cone
sample was lower than that of the \ycut\ and \ywin\ samples by a few
percent in very low and in high energy jets.  The \hw\ Monte Carlo
model and \jt\ in matrix element mode \cite{bib_jt74} predict similar
jet purities to the standard \jt\ samples, and detector effects on the
purity distribution were found to be small.

\subsection{Determination of Particle Rates}
\label{effi}

\begin{figure}[htb]
\begin{center}
\resizebox{\textwidth}{!}{\includegraphics{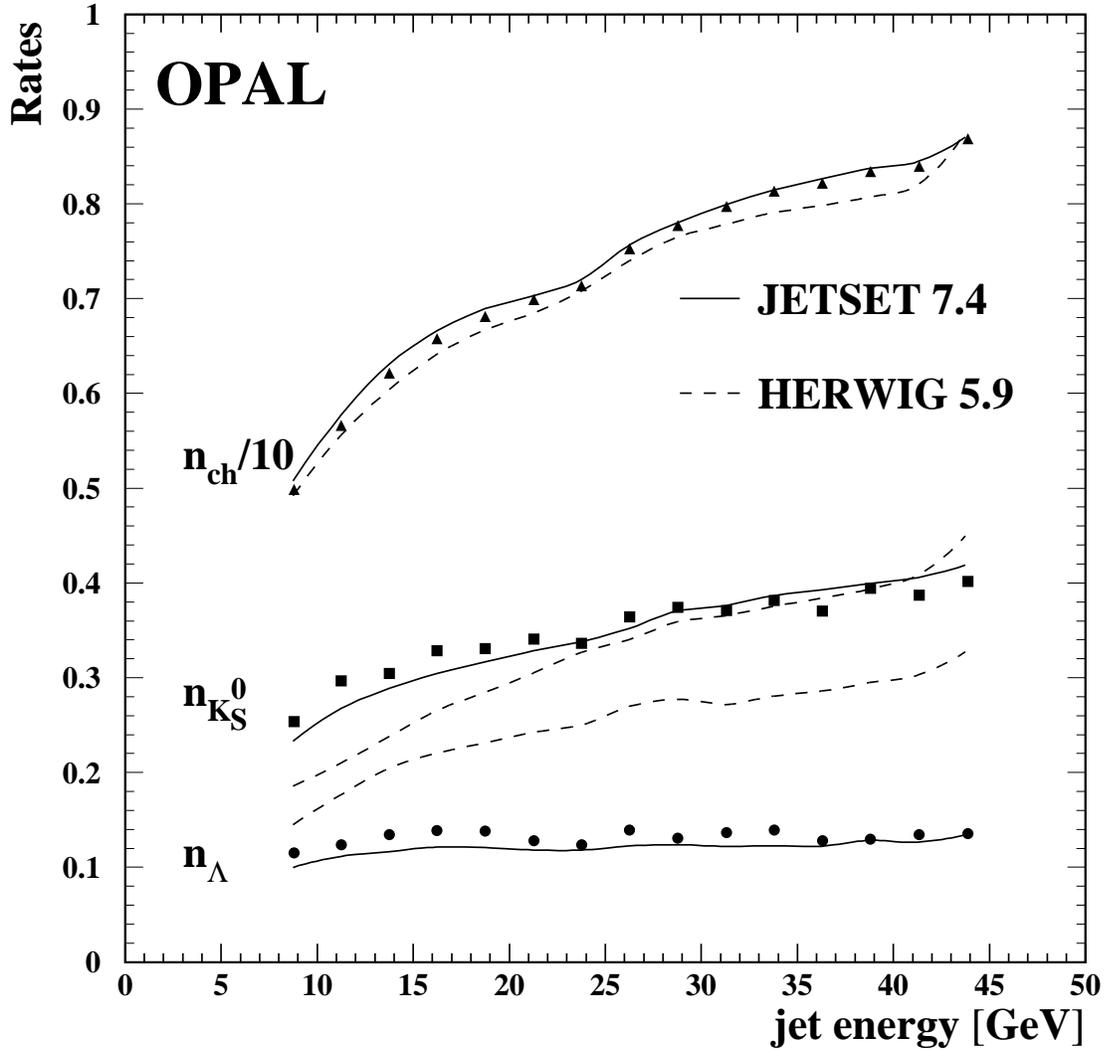}}
\end{center}
\caption
  {Production rates per jet of charged 
   particles, \ks\ mesons, and \lam\ baryons,
   $n_{\mathrm ch}/10$, $n_{K^0_S}$, and $n_{\Lambda}$, 
   from the \ycut\ sample as a function of the jet 
   energy compared with the predictions of the models \jt~7.4 and  
   \hw~5.9. The charged particle rates are scaled down 
   by a factor of 10. The errors shown are the (uncorrelated)
   statistical ones and are mostly smaller than the size of the symbols.}
\label{fig_b3}
\end{figure}

With the \ycut\ selection, about 100,000 \kzero\ and about 30,000
\lam\ were reconstructed.  The production rates per jet of charged
particles, \ks\ mesons and \lam\ baryons in jets, $n_{\mathrm {ch}}$,
$n_{{\mathrm K}^0_{\mathrm S}}$, and $n_{\Lambda}$, are shown as
functions of the jet energy in figure~\ref{fig_b3} after corrections
for reconstruction efficiency (about 16\%) and detector acceptance.
The decay products of \ks\ and \lam\ were not counted as charged
particles. The predictions of the \jt~7.4 and \hw~5.9 Monte Carlo
models are also shown in figure~\ref{fig_b3}. Whereas the \jt~7.4
generator describes the experimental data fairly well, the predictions
of \hw~5.9 agree poorly with the data. Similar results were obtained
with the {\tjet}s from the \ywin\ and the cone selections.

\subsection{Relative Rates and Correction for Jet Impurity}
\label{fits}

\begin{figure}[htb]
\begin{center}
\resizebox{\textwidth}{!}{\includegraphics{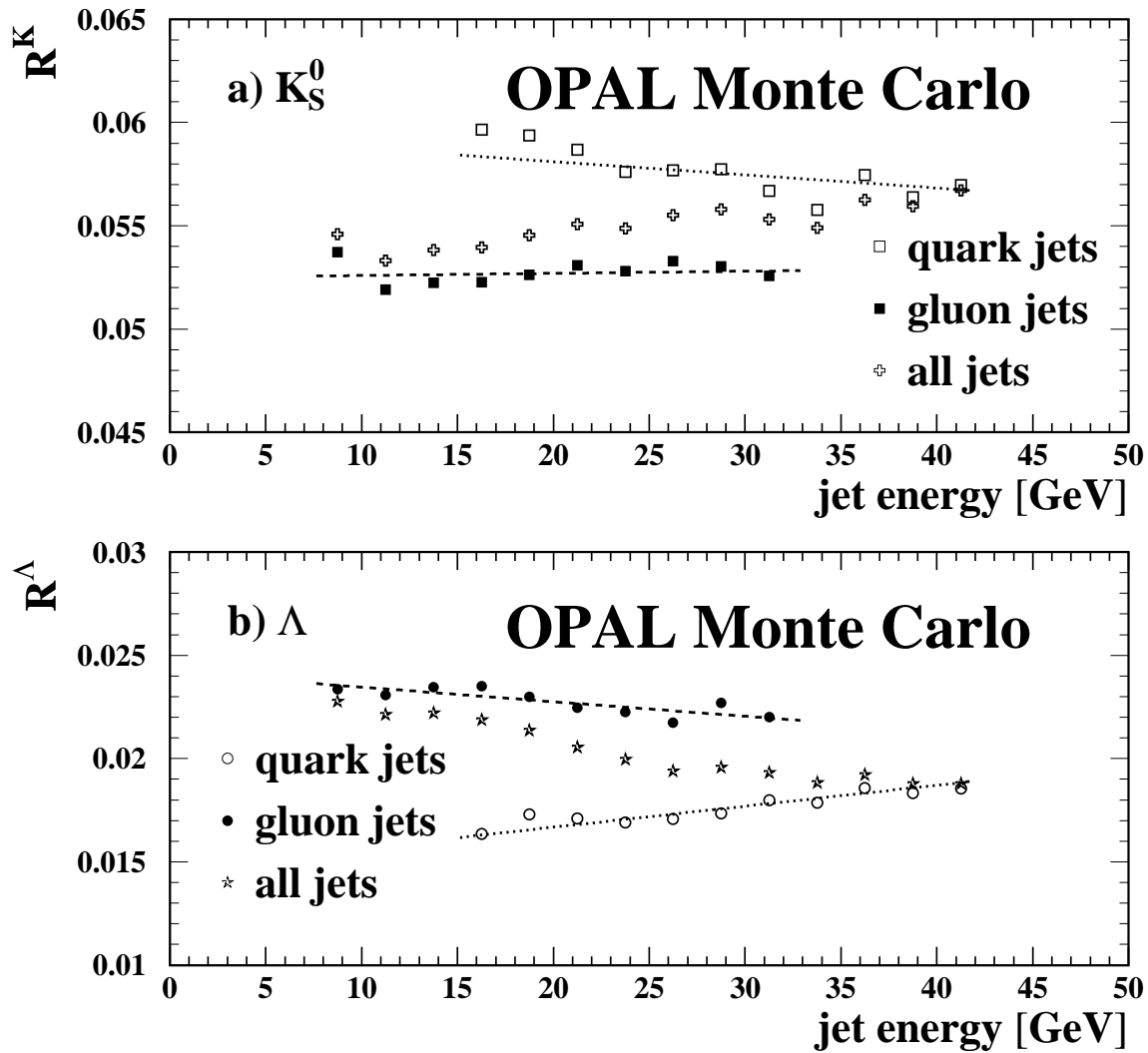}}
\end{center}
\caption
 {Relative production rates of (a) \ks\  and (b) \lam\ in \jt~7.4 events
  for pure quark and gluon jets as a function of the jet energy. The
  lines are fits of straight lines to the points. The statistical errors
  are smaller than the size of the symbols. The zeros of the vertical axes
  have been suppressed.}
\label{fig_relrates}
\end{figure}

The particle production rates rise with the jet energy partly due to
the changing mixture of gluon and quark jets, and partly due to the
increased energy available for particle production. In order to
measure differences due to quark and gluon jets, it is necessary to
remove this jet energy dependence.  If a similar energy dependence for
the production rates of all hadron species is assumed, then the
relative particle production rate (defined as the rate of \kzero\ or
\lam\ production divided by the rate of charged particle production,
$R_{{\mathrm K}^0_{\mathrm S}} =n_{{\mathrm K}^0_{\mathrm
    S}}/n_{\mathrm {ch}}$ and $R_{\Lambda}=n_{\Lambda}/n_{\mathrm
  {ch}}$) would be independent of the jet energy.  Studies of \jt\ 
events showed that there is indeed only a weak energy dependence of
these relative production rates in pure quark and gluon jets,
figure~\ref{fig_relrates}.  The lines are fits of straight lines and
have slopes smaller than $2 \times 10^{-4}~{\rm GeV}^{-1}$.  For gluon
jets the slopes are smaller than for quark jets, and for \ks\ smaller
than for \lam . As \jt\ gives a good overall description of the data
over a large range of c.m. and jet energies, it was assumed for
further analysis that the relative particle production rates are
constant in the data and we choose this assumption to reduce the
dependence on Monte Carlo models.  The consistency of the results with
this assumption will be shown later.  Particle production in jets
depends not only on the jet energy but also on the proximity of the
other jets in the event\cite{aleph_topo}.  Further Monte Carlo studies
showed that this angular dependence is almost the same for all hadron
species and does not affect the relative rates.


The relative particle production rates of \kzero\ and \lam\ were
computed in each jet energy interval from the ratios of the rates
shown in figure~\ref{fig_b3}. The relative rates for pure quark and
gluon jets were unfolded using the jet purities for each \tjet\ 
sample, obtained from Monte Carlo results shown in figure~\ref{fig_b2}.
A fit was performed to the observed relative rates as a function of
jet energy $E_j$ using the function:
\begin{equation}
R^h_m(E_j) = R^h_g \cdot \rho_g(E_j) + R^h_q \cdot (1 - \rho_g(E_j)) ,
\label{eqn_fit}
\end{equation}
where $R^h_m(E_j)$ is the measured relative particle rate, $\rho_g(E_j)$
the gluon purity of the jets, and $R^h_q$ and $R^h_g$ the relative rates
in pure quark and gluon jets, with $ h = \kzero , \lam ~$($h$ = hadron).
$R^h_q$ and $R^h_g$ were assumed to be constant for the reasons 
stated above. 

\begin{figure}[htb]
\begin{center}
\resizebox{\textwidth}{!}{\includegraphics{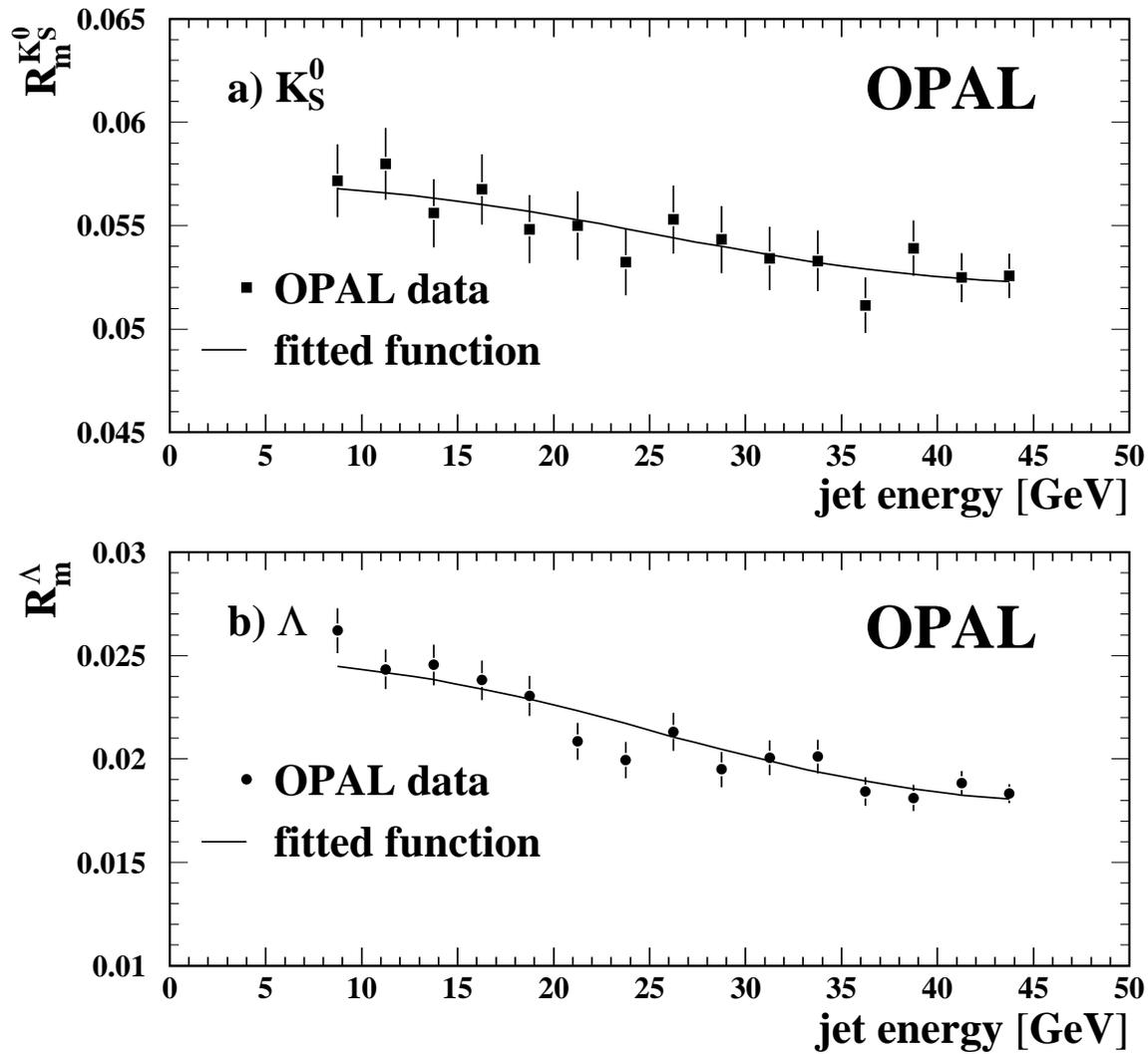}}
\end{center}
\caption
 {Relative production rates of (a) \ks\ and (b) \lam\ from the \ycut\ 
  selection as a function of the jet energy.  The lines 
  show the functions returned from the fits of equation~\ref{eqn_fit} to the
  data. The zeros of the vertical axes have been suppressed, and the
  errors shown include both statistical and systematic uncertainties.}
\label{fig_b4}
\end{figure}

The fit was performed in the jet energy range from 7.5~GeV to
45.0~GeV. The relative particle rates as a function of the jet energy
are shown in figure~\ref{fig_b4} for the \ycut\ selection, with the
lines giving the fit results. The fit results and $\chi^2$/d.o.f.
values for all three selections are given in
table~\ref{table_eoresults}. In order to compare particle production
in quark and gluon jets, the ratios $(R_g/R_q)_{\Lambda}$ and
$(R_g/R_q)_{\rm K}$ were studied.  These are given in
table~\ref{tab_b3} and will be discussed fully in
section~\ref{discuss}.

\subsection{Systematic Uncertainties}
\label{systematics}

The following sources of systematic uncertainty on the measured ratios
of relative rates have been studied.  For each source of uncertainty,
the difference with respect to the standard analysis was used to
estimate a symmetric systematic uncertainty. The uncertainties listed
in table~\ref{tab_b2} were added in quadrature to arrive at a total
systematic error.

\begin{description}
\item[Three-jet event reconstruction:] The analysis was repeated with
  the following changes:
\begin{itemize}
\item charged tracks only were used for the reconstruction of jets,
  instead of charged tracks and unassociated electromagnetic clusters.
  This check disregards all calorimeter information when determining
  energy flows in the events, so the changes represent an extreme
  situation. A systematic error was therefore determined from the
  difference divided by $\sqrt{12}/2\; (=1.7)$;
  
\item the resolution parameter of the \durham\ jet-finder was varied
  by replacing the \ycut\ = 0.005 selection by selecting
  three-jet events
  in a window of y-values (0.004, 0.008). The \ywin\ selection was modified
  by requiring a cut at a fixed value, $\ycut\ = 0.01$.

\end{itemize}

\item[\lam\ and \kzero\ reconstruction:] As in \cite{PR126} and
  \cite{PR061} the major sources of systematic uncertainties for the
  reconstruction of \ks\ and \lam\ were found to be the background
  determination and the reliability of the Monte Carlo simulation for 
  distributions on which cuts were placed. 

\begin{itemize}
\item The \kzero\ and \lam\ selection criteria were varied as in
  \cite{PR126} and \cite{PR061}. In particular, the  cut on the
  distance between the reconstructed secondary vertex and the first
  measured hit of the decay particles was loosened from 3~to 9~cm;
\item a sideband method~\cite{PR061} was applied to determine the backgrounds 
  under the \ks\ and \lam\ signals; and
\item systematic uncertainties on the strange particle reconstruction
  efficiencies were estimated by calculating the efficiencies
  separately using the \jt~7.3 and \jt~7.4 samples (the
  standard analysis used the combined \jt~7.3 and 7.4 samples). The
  simulation of baryon production differs considerably in these samples
  and they were taken to represent alternative possibilities of
  generator tuning and detector simulation.  Studies of events
  generated using \hw\ as input to the detector simulation gave
  efficiencies lying in the range spanned by \jt~7.3 and \jt~7.4. A
  symmetric systematic uncertainty was assigned using the full difference
  between \jt~7.3 and \jt~7.4 divided by  $\sqrt{12}$.
\end{itemize}
\item[Quark and gluon jet unfolding:]
The systematic uncertainties in the determination of \ks\ and \lam\ production
in quark and gluon jets were obtained by the following variations:

\begin{itemize}
\item the upper and lower bounds of the fit range ($7.5 - 45.0$ GeV)
  were changed to 12.5 and 40~GeV respectively; 
  
\item the influence of the jet purity determination was studied by
  varying the purities by their systematic uncertainties (about 5\%),
  which were derived as described in~\cite{PN236}.  Briefly, the
  uncertainty in the identification of quark and gluon jets was
  estimated by comparing different fragmentation models (\jt\ and
  \hw) and studying detector resolution effects;
  
\item as a check of the fit procedure, the relative production rates
  in pure gluon jets were obtained by a linear fit to the relative
  rates $n_m(\rho_g$) as a function of the gluon jet purity $\rho_g$
  above $\rho_g \geq$ 8\%, and extrapolating to 100\% gluon jet purity;
  
\item the fit function was modified to account for a possible linear
  energy dependence of the relative particle production rates in pure
  quark and gluon jets, using the \jt\ slopes as shown in
  figure~\ref{fig_relrates}.  It can be seen from table~\ref{tab_b2} 
  that considering non-zero values of the slopes results in only a small
  contribution to the systematic uncertainties. This means that the relative
  rates found in the experimental data are consistent with the assumption
  of independence of the jet energy.
\end{itemize}

\end{description}

The main contributions to the systematic uncertainties came from the
variation of the cuts on \ks\ and \lam\ candidates, from the
differences in detection efficiencies determined using \jt~7.3 and
\jt~7.4, and from the variation of the fit ranges.  The ratios of
relative rates determined from the fit were essentially unchanged if
contributions from \ks\ and \lam\ decay products were included in the
charged particle rates.

Finally, Monte Carlo events with full detector simulation were
analysed in the same manner as the data. The relative rates derived
from the fits (0.95$\pm$0.02 for \kzero\ and 1.27$\pm$0.04 for \lam)
were compared with those determined directly at the generator level in
these events (0.94 and 1.26 respectively).  The agreement for both
\lam\ and \ks\ from the \ycut\ sample is good. The results from the
other two jet samples agreed equally well.

\section{The Y-event Analysis}\label{Yanalysis}

\subsection{Three-jet Event Selection}
\label{selection}

For the Y-event analysis, jets were defined with  the cone 
jet finding algorithm, supplying all accepted particles as 
input.  The resolution parameters chosen for the jet finder were a
cone size $R=30\degs$, as in an earlier publication~\cite{PR141}, and
a minimum jet energy \eps\ computed once for each event according to
$\eps=5\times\evis/\roots$~GeV where \roots\ is the c.m. energy and
\evis\ the sum of the energy of the particles. 

The criteria given in references~\cite{PR038,PR070,PR136,PR141} were
followed to select a sample of symmetric three-jet events.  Each jet
was required to contain at least two particles and to lie in the polar
angle region $|\cos\theta|<0.9$, and the sum of the angles between the
three jets was required to exceed $358^\circ$. As in the energy-based
approach, the jets in each event were assigned a calculated energy
based on the interjet angles, assuming massless kinematics.  Strongly
symmetric three-jet events were selected by projecting the jets into
the three-jet event plane and requiring the angles between the jet
with the highest calculated energy and each of the two others to be in
the range $150\degs \pm 10\degs$.  The event plane is defined as the
plane perpendicular to the sphericity \cite{sphericity} eigenvector
associated with the smallest eigenvalue.  In total, 70,738 symmetric
three-jet events were found. The mean calculated jet energies were
$42.50 \pm 0.01$~GeV for the highest energy jet and $24.37 \pm
0.01$~GeV for the two lower energy jets.  The highest energy jets are
likely to be quark jets with high probability, due to the nature of
the gluon radiation spectrum. From a Monte Carlo study, this
probability was estimated to be 97.3\%.  The two lower energy jets
were therefore assumed to be a quark jet and a gluon jet of equal
energy with the same angles with respect to the other two jets in the
event. The inclusive sample of lower energy jets is referred to as the
`normal-mixture' sample of jets.

\subsection{Gluon Jet Identification} 
\label{tag}

A gluon-jet enriched sample of the lower energy jets was selected by a
new variant of the method \cite{PR141,PR136,PR070,PR038} of
identifying quark jets in order to `anti-tag' the gluon jets.  For
each lower energy jet identified as a quark jet, the other lower
energy jet was assumed to be a gluon jet. This anti-tagged sample of
gluon jets was therefore essentially unbiased by the tagging method.
It contained a well known fraction of gluon jets established from
studies of simulated events.

Previous studies employed the reconstruction of secondary vertices or
the identification of energetic leptons to identify jets that
originated from heavy quarks. Whilst this yielded an anti-tagged jet
sample with a gluon purity typically in excess of 90\%, the efficiency
to identify quark jets was only of order 5\%. The present analysis
introduces a new method to isolate a large sample of quark jets by
identifying jets that are collimated, i.e. that have a large fraction
of their energy close to the jet axis. Studies of quark and gluon jets
in symmetric events have shown that gluon jets are broader than quark
jets~\cite{PR136} and this is well reproduced by Monte Carlo models.

\begin{figure}[htb]
\begin{center}
\resizebox{\textwidth}{!}{\includegraphics{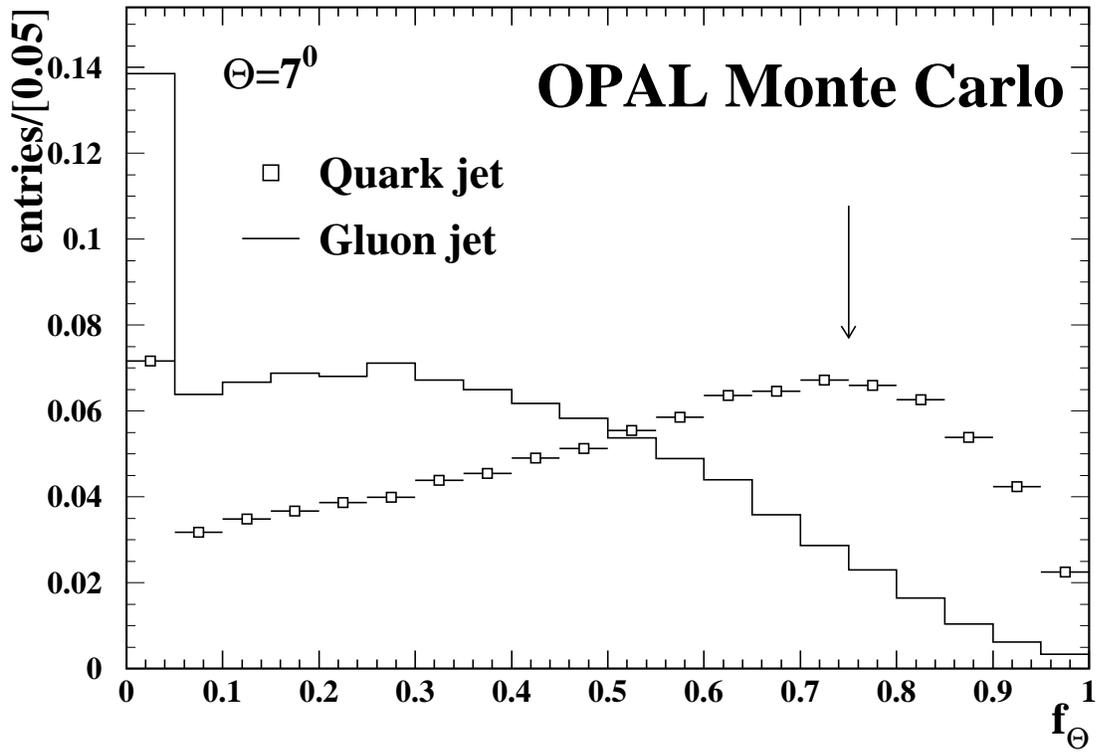}}
\end{center}
\caption
{ Distributions of \fthet\  for quark and gluon jets.  The events were
  generated with the \jt\ 
Monte  Carlo and include  full simulation of the detector. The arrow
  indicates the cut position for selecting quark jets.}
\label{fthet}
\end{figure}

In particular, the quantity \fthet\ is determined, which is given by
the fraction of the jet's energy contained in a cone co-axial with the
jet axis and with half-angle $\Theta$.  If the energy of the particles
contained in the sub-cone is $E_\Theta$ and the visible jet energy is
$E_\mathrm{jet}$ then $\fthet =\frac{E_\Theta}{E_\mathrm{jet}}$.  The
\fthet\ distributions were studied in detector level Monte Carlo
events.  Three-jet events were selected as above, and \fthet\ 
determined for the two lower energy jets.  Each simulated hadron jet
was associated with an underlying quark or gluon jet using the method
described in~\cite{PR136}.   Briefly, the two hadron jets that were closest
in angle to the directions of the primary quark and anti-quark which
had evolved from the \zzero\ decay were considered to be the quark
jets, and the remaining jet was identified as the gluon jet.
Figure~\ref{fthet} shows the distributions
of \fthet\ for $\Theta = 7\degs$ for jets in the simulated events that
were classified as quark or gluon jets.  The quark jets tend to have
larger \fthet\ values than the gluon jets, and
a sample of jets with a high quark content can therefore be selected
by requiring \fthet\ to exceed some threshold \fcut, -- for example
$\fcut\ = 0.75$.  

The anti-tagged jet purity \pur\ is defined to be the fraction of
anti-tagged jets that are indeed associated with an underlying gluon
jet, and the tagging rate, \rtag, is defined to be the fraction of
normal-mixture Y-events which contain jets that are anti-tagged.  From
Monte Carlo studies, the normal-mixture sample of jets had a gluon
content, $\pnmix = 48.7 \pm 0.2$\%.  The anti-tagged jet purity and
the tag rate depend on the values of $\Theta$ and \fcut, and values of
\pur\ of up to about 80\% can be achieved whilst maintaining a tag
rate in excess of about 30\%.  A high gluon purity is desirable for
the algebraic correction procedure later, and therefore
$\Theta=7\degs$ and $\fcut\ = 0.75$ were chosen for the standard tag.
From studies of Monte Carlo events the values $\rtag = 30.3 \pm 0.2\%$
and $\pur\ = 77.0 \pm 0.5 \pm 0.9\%$ were determined, where the first
error is statistical and the second systematic.  The systematic error
includes contributions from the choice of Monte Carlo model (\jt~7.4,
\jt~7.3 or \hw~5.9) and the method used to identify the quark jet in
the simulated events (the method described in section~\ref{purities}
for the energy-based analysis was used as an alternative).  In the
data, $\Theta=7\degs$ and $\fcut\ = 0.75$ gave a tag rate of $31.0 \pm
0.2\%$ (23,256 anti-tagged jets) which is well described by the
simulated events. The average energy of the anti-tagged jets was
$23.59 \pm 0.02$~GeV.

In about $1.9\%$ of events both of the lower energy jets fulfilled the
tag criteria due to misidentification of collimated gluon jets (the
highest energy jet is the gluon in fewer than 10\% of these cases),
and therefore in these events both the lower energy jets were included
in the anti-tagged gluon jet sample. The double tag rate rose to about
$3.5\%$ for $\Theta=8\degs$ and $\fcut\ = 0.75$.

The flavour composition of the tagged, anti-tagged and normal-mixture
jets in the Monte Carlo events is shown in table~\ref{tag_flavour}.
Significantly fewer b-flavour jets were tagged than light-flavour jets
which is consistent with the observation that b quark jets in \zzero\ 
decays are broader than light quark jets \cite{PR141}; a small
reduction is also visible in the rate of tagging c-flavour jets. The
light quark flavour composition of the tagged jets reflects the
relative couplings of u- and d-type quarks to the \zzero. There is a
fairly good correspondence between the quark flavour properties of the
normal-mixture and anti-tagged jet samples. It was shown
in~\cite{PR136} that there is no significant systematic bias to the
measurements of quark and gluon jet differences from the flavour
composition of the anti-tagged jets. The flavour composition of the
quark jets in the anti-tagged sample is therefore not expected to give
a systematic bias in the present analysis.

\subsection{Correction Methods}
\label{corrections}

\subsubsection*{Strange Particle Identification}
\label{finders}

\kzero\ mesons and \lam\ baryons were reconstructed as described in
section~\ref{lamk0s}. Their invariant mass spectra were computed in
bins of momentum, \pkzero\ and \plam, within the normal-mixture and
anti-tagged jets separately. A \kzero\ or a \lam\ candidate was
assigned to a jet if its momentum vector fell within the cone defining
the jet, i.e., if it lay less than 30\degs\ from the jet axis.
If a \kzero\ or \lam\ was within 30\degs\ of both jets, it 
was assigned to the higher (calculated) energy jet.

The detection efficiencies of the \kzero\ and \lam\ finding algorithms
were sensitive to the particle momenta and to the track environment,
and were therefore determined from simulated events for each bin of
\pkzero\ and \plam\ within the normal-mixture and anti-tagged jets
separately.  The efficiency was computed from the fraction of
generated \kzero\ or \lam\ within a jet that were found in the same
jet by the algorithm.  The background-subtracted numbers of \kzero\ 
and \lam\ were corrected for detection efficiency as a function of
momentum, and summed to give the production rates of \kzero\ and \lam\ 
in normal-mixture and anti-tagged jets, $D^{\kzero}_{\mathrm{n.mix}}$
and $D^{\kzero}_{\mathrm{a.tag}}$, $D^{\Lambda}_{\mathrm{n.mix}}$ and
$D^{\Lambda}_{\mathrm{a.tag}}$.

\subsubsection*{Algebraic Correction Procedure}

The algebraic method introduced
in~\cite{PR070} was used to correct for quark and gluon
misidentification and to arrive at the ratio of the identified
particle production rates for pure quark and gluon jets.

The production rate of  particle type $i$  in the normal-mixture sample of
jets, $D^i_{\mathrm{n.mix}}$  may be written 
\begin{equation}
\label{dnmix}
 D^i_{\mathrm{n.mix}} = \pnmix\cdot G^i\;
                 + \;(1-\pnmix) \cdot Q^i\, \;\; ,
\end{equation}
where $G^i$ and $Q^i$  are the production rates of $i$ in pure gluon and
quark  jets respectively. Similarly, the production rate in the anti-tagged
sample  $D^i_{\mathrm{a.tag}}$ may be written 
\begin{equation}
\label{dtag}
 D^i_{\mathrm{a.tag}} = \pur\cdot G^i\;
                 + \;(1-\pur) \cdot Q^i\, \;\; .
\end{equation}
 
The ratio of the production rates for pure gluon and quark jets \rgqi\
may therefore be determined by
\begin{equation}
\label{algebraic}
   \rgqi = \frac{ G^i}{Q^i} = \frac
{ (1-\pnmix)\cdot(D^i_{\mathrm{a.tag}}/D^i_{\mathrm{n.mix}}) -(1-
\pur) } 
{\pur - \pnmix(D^i_{\mathrm{a.tag}}/D^i_{\mathrm{n.mix}})} \;\; .
\end{equation}

As in the case of the multiplicity measurement in~\cite{PR141} it was
assumed for the purposes of the algebraic correction that $Q^i$ and
$G^i$ are the same in equations \ref{dnmix} and \ref{dtag}. This is a
reasonable assumption, as the quark flavour compositions of the
normal-mixture and anti-tagged samples are generally consistent, and
as the properties of energetic acolinear gluons are independent of
event flavour according to QCD. The simulated events provide a good
representation of the relevant event properties such as the
collimation of jets.  Therefore any residual effects are expected to
be removed by the corrections for detector effects.
     

The measured production rates $D^i$ appear in equation~\ref{algebraic}
as the ratio $(D^i_{\mathrm{a.tag}}/D^i_{\mathrm{n.mix}})$ and therefore
some systematic effects related to particle identification are
expected to cancel.  Statistical uncertainties on \rgqi\ were
estimated from the variance of the results obtained when the analysis
was repeated ten times with the data split into independent subsets.
This procedure correctly takes into account correlations between
$D^i_{\mathrm{a.tag}}$ and $D^i_{\mathrm{n.mix}}$.

\subsubsection*{Detector Corrections}

A correction derived from Monte Carlo events was applied to correct
for detector acceptance and resolution.  The correction was formed
from the ratio of \rgqi\ values determined at the generator and
detector levels.  At the generator level, the same three-jet event
selection criteria as for the data were applied (with the exception of
the requirement $|\cos\theta|<0.9$ for the jet axes). Monte Carlo
information was used to identify quark and gluon jets, as well as
particle type.  The detector corrections were determined to be $1.105
\pm 0.005$ for charged particles, $1.023 \pm 0.006$ for \kzero\ and
$1.058 \pm 0.008$ for \lam\ with uncertainties due to the limited
statistics of the Monte Carlo samples.  The correction factor for
charged particles agrees well with that determined in \cite{PR136}.
The final results can be found in table~\ref{T_summary} which will be
discussed later.

\subsection{Systematic Uncertainties}
\label{syst}

Systematic uncertainties, which are listed in table~\ref{T_syst},
were evaluated from a number of sources in a
similar fashion to the energy-based analysis described in
section~\ref{systematics}. The
total systematic errors were obtained by combining the individual
contributions in quadrature.

\begin{description}
  
\item[Detector effects:] The total energy and momentum flow in the
  event were estimated using charged tracks only instead of tracks and
  electromagnetic clusters.  These were used as input to the jet
  finding and quark jet identification algorithms and the analysis was
  repeated.  The observed jets are more collimated when measured using
  tracking information only. This is reproduced by the Monte Carlo,
  both at the generator and detector levels and is partly due to
  decays of $\pi^0$ into photons which tend to broaden the flow of
  neutral particles in the jet.  The effect of this increased
  collimation was that the anti-tagged jet purities and tag rates were
  somewhat different to the standard case, with $\pur = 70.2 \pm
  0.4$\% and $\rtag\ = 42.0 \pm 0.2$\% for $\Theta=7\degs$ and $\fcut\ 
  = 0.75$. The uncertainty was estimated from the difference divided
  by $\sqrt{12}/2$ as for the energy-based analysis above.
  
\item[Quark jet identification:] The analysis was repeated using
  parameters $\Theta=8\degs$ and $\fcut\ = 0.75$ since its tagging
  rate ($\rtag = 40.1 \pm 0.2$), and the anti-tagged jet purity
  ($\pur\ = 74.7 \pm 0.4 \pm 0.9$) were somewhat different compared to
  the standard analysis.

\item [Event selection:] Events were selected requiring a minimum jet
  energy of 10~GeV which is the selection criterion 
  used in~\cite{PR136}.  
  
\item [Jet purity determination:] A number of sources of systematic
  uncertainty on \pur\ have been considered. Simulated events from the
  \jt~7.4 and \jt~7.3 samples were used separately to determine the
  purities and tag rates, and the observed differences were used to
  evaluate an uncertainty related to the event generator tuning and
  detector simulation.  A further uncertainty comes from the
  ambiguity in defining whether a jet arises from a quark or a gluon,
  as described in a previous OPAL publication \cite{PR141}. The same
  procedures were followed to determine a systematic error which was
  added in quadrature with the other sources to arrive at the total
  systematic uncertainty.  The analysis was repeated using the tagged
  and normal-mixture jet purity values varied by their combined
  statistical and systematic uncertainties.
  
\item[Background determination:] The fit ranges, signal window
  sizes and excluded regions were all varied 
  to determine the background to the
  selected \kzero\ and \lam\ signals. In the case of the \lam\ 
  an alternative function was also used to describe the background. 
  Finally a sideband method was used to determine the background.
    
\item[Efficiency determination:] The \jt~7.3 and \jt~7.4 samples
  of simulated events were used separately to estimate the efficiency of the
  \kzero\ and \lam\ finding algorithms. The two Monte Carlo samples
  are taken to represent alternative possibilities of generator tuning
  and detector simulation and  
  an uncertainty was estimated as for the energy-based analysis above.
  Several of the systematic variations were made simultaneously, 
  and in no case was a difference larger than that for
  \jt~7.3 observed. 

\item[Monte Carlo statistics:] The finite numbers of simulated events
  available led to statistical uncertainties in the particle
  detection efficiencies, and the detector corrections. 

\end{description}

The effect of varying the cone size $R$ that defines the jets has been
investigated in~\cite{PR136} where an increase in \rgqch\ with $R$ was
observed.  Other sources of uncertainty such as changing the
requirement on the angle between the highest energy and the other
jets, and modifying the track and cluster selection criteria have been
considered in that publication and found to be negligible.  There were
also no statistically significant differences between the detector
acceptance corrections computed with the \jt~7.3 or \jt~7.4 samples.

\section{Results and Discussion}
\label{discuss}

\label{eoresults}

The results of the energy-based analysis are given in
table~\ref{tab_b3}.  The ratios $R^{\rm K^0_S}_g/R^{\rm K^0_S}_q$ and
$R^{\Lambda}_g/R^{\Lambda}_q$ 
of relative \ks\ and \lam\ production rates in
pure gluon and quark jets are given for all three event selections,
together with their statistical and systematic uncertainties and the
predictions of the \jt~7.4 model.  For the \ycut\ selection, the
production of \ks\ mesons is enhanced in gluon jets relative to quark
jets by a factor
$1.10 \pm 0.02 \pm 0.02$ where the first error is statistical
and the second systematic. Similar results were obtained with the other
event selections.  The relative production rate of \lam\ baryons was
found to be increased in gluon jets relative to quark jets by $1.41
\pm 0.04 \pm 0.04$ in the \ycut\ selection.  The less collimated \ywin\ 
and cone event samples show a smaller increase in the relative
production rates of \lam\ baryons indicating a possible dependence of
the baryon production rates on the topology of the events.

For the Y-event analysis, the ratios of absolute production rates of
\kzero\ and \lam\ in 24~GeV gluon and quark jets were found to be
$\rgqk = 1.05 \pm 0.08 \pm 0.09$ and $\rgql = 1.32 \pm 0.11 \pm 0.18$
for jets defined with the cone algorithm using a cone size of 30\degs.
The corresponding result for charged particles is $\rgqch = 1.116 \pm
0.006 \pm 0.012$ which is in good agreement with the result from the
previous OPAL vertex-tagged analysis ($1.10 \pm 0.02 \pm 0.02$
\cite{PR136}) and which has a reduced statistical error as a result of
the use of the more efficient energy flow tag.

The measurements in the Y-event analysis may also be used to determine
production rates of \kzero\ and \lam\ in quark and gluon jets relative
to those of charged particles.  The relative rates of \kzero\ and
\lam\ production are $R^{\rm K^0_S}_g/R^{\rm K^0_S}_q = 0.94 \pm 0.07
\pm 0.07$ and $R^{\Lambda}_g/R^{\Lambda}_q = 1.18 \pm 0.10 \pm 0.17$,
where correlations between the sources of systematic uncertainty
(given in table~\ref{T_syst}) have been taken into account.

An enhancement of \lam\ production in gluon jets relative to quark
jets, in excess of that observed for charged particles, is measured by
both analyses, with the ratios of the relative \lam\ rates consistent
within the errors. The ratios of the relative production rates of
\kzero\ mesons in gluon and quark jets are also compatible within the
errors, and suggest a small enhancement relative to charged particles.
The analyses presented here are consistent with previous findings
\cite{PR136,aleph_topo} that have shown that factors such as jet
finder, jet energy and event topology are important in quantifying the
differences between quark and gluon jets. Therefore, care should be
taken to ensure that the conditions are equivalent when comparing
results between experiments.

\begin{figure}[htb]
\begin{center}
\resizebox{\textwidth}{!}{\includegraphics{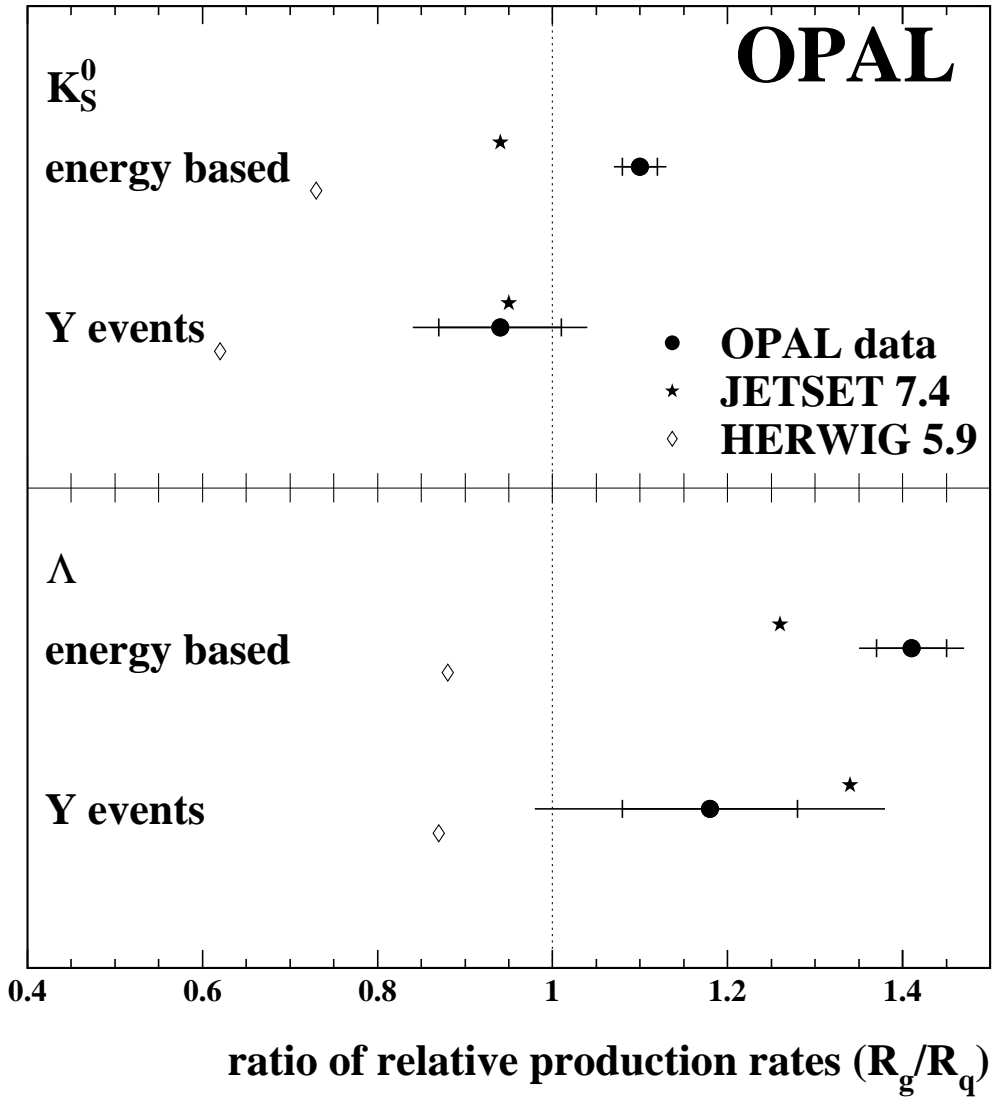}}
\end{center}
\caption
{The ratio of relative production rates (see text) in quark and gluon
  jets of \kzero\ and \lam\ for both analyses.  The experimental
  statistical errors are delimited by the small vertical bars. The
  predictions of \jt~7.4 and \hw~5.9 are also shown.  The predictions
  of \jt~7.3 are no more than 0.04 lower than those of \jt~7.4.}
\label{F-summary}
\end{figure}

The measurements of relative production rates in gluon and quark jets
from both analyses are shown in figure~\ref{F-summary} together with
the predictions of the \jt~7.4 and \hw~5.9 Monte Carlo models. These
data are also given in table~\ref{T_summary} together with the
measurements of absolute rates from the Y-event analysis.  \hw~5.9,
despite its good description of global event properties, fails to give
an adequate description of the measurements of ratios of strange
particle production rates in quark and gluon jets. This result is not
surprising given the poor description of the inclusive strange
particle rates as a function of jet energy shown in
figure~\ref{fig_b3}.  \jt~7.4, however, was shown to provide a
reasonable general description of the data in figure~\ref{fig_b3}, and
the ratios of relative rates from the Y-event sample are consistent
with its predictions for both \kzero\ and \lam.  The ratios of
relative rates from the energy-based analysis, however, are
significantly larger than those predicted for both \kzero\ and \lam.
The predictions of the ratios of relative production rates from
\jt~7.3 differ from \jt~7.4 by at most 0.04 which gives an indication
of the size of possible effects due to parameter tuning and inclusion
of additional particle decay channels.

There is no perturbative mechanism in the \jt\ model that gives rise
to the observed differences in particle production between quark and
gluon jets; they arise from the effects of hadronization and particle
decays. Many of the \kzero\ are the decay products of heavy (b and c
flavour) hadrons. As the production of b and c quarks in gluon jets is
suppressed, there are correspondingly fewer \kzero\ in these jets.
The enhancement of \lam\ production (and of baryons in general) in gluon
jets relative to quark jets predicted by \jt\ is a consequence of
the different dynamics of the string fragmentation process in quark
and gluon jets.

\section{Summary}
\label{summary}

Production rates for \ks\ and \lam\ have been measured in quark and gluon
jets from \zzero\ decays with two complementary approaches.  In the
first analysis a procedure was introduced to compare particle
production in gluon and quark jets of different energies facilitating
the study of up to about $24$\% of the total event sample. Different
jet finding algorithms (\durham, cone) were used to investigate
samples with different three-jet event topologies. Relative rates,
normalized to the inclusive charged particle rate, were obtained for
pure quark and gluon jets considering the different quark and gluon
content of jets in different energy intervals. The relative rates in
pure gluon and quark jets were found to be consistent with being
independent of the jet energies, and to depend slightly on the
specific jet selection.  

In the second analysis a new method was introduced to tag quark jets
based on the collimation of their energy flow, allowing the isolation
of a larger sample of anti-tagged gluon jets in symmetric three-jet
events than the method of secondary vertex tagging used previously.
The comparison of quark and gluon jets of equal energies and embedded
in almost identical event environments allows for a simple interpretation of
the results. The jets were selected using a cone algorithm.
By also measuring the inclusive particle rates in these symmetric jets,
relative rates were obtained in addition to the absolute rates.

An enhancement of \lam\ production in gluon jets relative to quark
jets, in excess of that for charged particles is observed.  The ratios
of production rates of \kzero\ mesons in gluon and quark jets suggest
a small enhancement relative to charged particles. The results of both
analyses are compatible within their errors. 
The predictions of \jt~7.4 are consistent with the enhancement
observed for \lam, but are smaller for the \kzero, whilst \hw\ fails
to provide an adequate description of the data.

\bigskip \bigskip
\noindent
{\bf Acknowledgements }

We particularly wish to thank the SL Division for the efficient operation
of the LEP accelerator at all energies
 and for their continuing close cooperation with
our experimental group.  We thank our colleagues from CEA, DAPNIA/SPP,
CE-Saclay for their efforts over the years on the time-of-flight and trigger
systems which we continue to use.  In addition to the support staff at our own
institutions we are pleased to acknowledge the  \\
Department of Energy, USA, \\
National Science Foundation, USA, \\
Particle Physics and Astronomy Research Council, UK, \\
Natural Sciences and Engineering Research Council, Canada, \\
Israel Science Foundation, administered by the Israel
Academy of Science and Humanities, \\
Minerva Gesellschaft, \\
Benoziyo Center for High Energy Physics,\\
Japanese Ministry of Education, Science and Culture (the
Monbusho) and a grant under the Monbusho International
Science Research Program,\\
German Israeli Bi-national Science Foundation (GIF), \\
Bundesministerium f\"ur Bildung, Wissenschaft,
Forschung und Technologie, Germany, \\
National Research Council of Canada, \\
Research Corporation, USA,\\
Hungarian Foundation for Scientific Research, OTKA T-016660, 
T023793 and OTKA F-023259.\\

\clearpage
\section*{Tables}


%
\begin{table}[htb]
\begin{center}
\begin{tabular}{|c|c|c|c|}
  \hline
Energy-based & $R^{\rm K^0_S}_g$ & $R^{\rm K^0_S}_q$ & $\chi^2$/d.o.f. \\
\hline
\ycut\ & $ 0.0573 \pm 0.0009$ & $ 0.0522 \pm 0.0006 $ &  9/13 \\
\ywin\ & $ 0.0568 \pm 0.0009$ & $ 0.0531 \pm 0.0006 $ &  8/13 \\
cone   & $ 0.0622 \pm 0.0011$ & $ 0.0580 \pm 0.0007 $ & 14/13 \\
\hline
\hline
Energy-based & $R^{\Lambda}_g$   & $R^{\Lambda}_q$   & $\chi^2$/d.o.f. \\
\hline
\ycut\ & $ 0.0252 \pm 0.0005$ & $ 0.0179 \pm 0.0003 $ & 15/13 \\
\ywin\ & $ 0.0252 \pm 0.0005$ & $ 0.0188 \pm 0.0003 $ & 12/13 \\
cone &   $ 0.0281 \pm 0.0006$ & $ 0.0214 \pm 0.0004 $ & 39/13 \\
\hline
\end{tabular}
\end{center}
\caption{\protect Fitted relative \ks\ and \lam\ production rates in 
  gluon and quark jets from the different event selections of the
  energy-based analysis. The results have been obtained by a fit to
  equation 1; the fit quality is indicated by the $\chi^2$/d.o.f.
  values. The errors are statistical.}
\label{table_eoresults}
\end{table}

\begin{table}[htb]
\begin{center}
\begin{tabular}{|c|c|c|}
  \hline
Energy-based & $R^{\rm K^0_S}_g/R^{\rm K^0_S}_q$ & \jt~7.4 \\
\hline
\ycut\ & $ 1.10 \pm 0.02 \pm 0.02 $  & 0.94 \\
\ywin\ & $ 1.07 \pm 0.02 \pm 0.02 $  & 0.94 \\
cone &   $ 1.07 \pm 0.02 \pm 0.02 $  & 0.95 \\
\hline
\hline
Energy-based & $R^{\Lambda}_g/R^{\Lambda}_q$ & \jt~7.4 \\
\hline
\ycut\ & $ 1.41 \pm 0.04 \pm 0.04 $  & 1.26 \\
\ywin\ & $ 1.34 \pm 0.04 \pm 0.03 $  & 1.24 \\
cone &   $ 1.31 \pm 0.04 \pm 0.05 $  & 1.30 \\
\hline
\end{tabular}
\end{center}
\caption{\protect\label{tab_b3}Ratios of the relative \ks\ and \lam\ 
  production rates in gluon and quark jets from the different event
  selections of the energy-based analysis compared to the \jt~7.4 predictions.
  The first error is statistical and the second systematic.}
\label{results_EB}
\end{table}

\begin{table}[htb]
\begin{center}
\begin{tabular}{|l|ccc|ccc|} \hline 
Source of error & \multicolumn{3}{c|}{\ks} 
                & \multicolumn{3}{c|}{\lam} \\ \hline \hline
Statistical error      & 2.1 \% & 2.3 \% & 2.3 \% & 2.7 \% & 3.0 \% & 3.2 \% \\
\hline
Charged tracks only      & & 0.1 \% & & & 0.4 \% & \\ 
\ycut /\ywin\ change     & & 0.2 \% & & & 0.2 \% & \\ \hline 
\lam /\ks\ cut variation & & 1.1 \% & & & 1.7 \% & \\ 
Sideband fit             & & 0.6 \% & & & 0.1 \% & \\ 
JETSET 7.3/7.4           & & 0.4 \% & & & 0.9 \% & \\  \hline
Fit range              & 0.7 \% & 0.7 \% & 0.9 \% & 1.4 \% & 0.5 \% & 1.4 \% \\
Jet purities           & 0.3 \% & 0.3 \% & 0.4 \% & 0.8 \% & 0.4 \% & 0.5 \% \\
Fit method             & 0.6 \% & 0.7 \% & 0.5 \% & 0.7 \% & 0.8 \% & 2.2 \% \\
MC slopes              & 0.3 \% & 0.2 \% & 0.8 \% & 0.7 \% & 0.7 \% & 1.2 \% \\
\hline 
Total systematic error & 1.7 \% & 1.8 \% & 2.0 \% & 2.7 \% & 2.3 \% & 3.5 \% \\
\hline
\end{tabular}
\caption[Statistical and systematic errors of particle ratios]
{\protect\label{tab_b2}Statistical and systematic errors of the ratios
  of relative particle production rates in gluon and quark jets in the
  energy-based analysis. The three numbers in a row refer to the
  \ycut, \ywin, and cone selection.}
\end{center}
\end{table}

\begin{table}[htb]
\begin{center}
\begin{tabular}{|c|c|c|c|}
  \hline
  & Anti-tagged & Tagged & Normal-mixture \\
\hline
d &  $5.1 \pm 0.1$ & $ 22.1\pm 0.2$ & $11.3\pm 0.1$ \\
u &  $4.2 \pm 0.1$ & $18.4\pm 0.2$ & $9.1 \pm 0.1$ \\
s &  $5.2 \pm 0.1$ & $23.4\pm 0.2$ & $11.4\pm 0.1$ \\
c &  $4.2 \pm 0.1$ & $12.8\pm 0.1$ & $ 9.0\pm 0.1$ \\
b &  $4.2 \pm 0.1$ & $4.3 \pm 0.1$ & $10.5\pm 0.1$ \\
gluon &  $77.0 \pm 0.5$ & $19.0\pm0.2$ & $48.7\pm 0.2$ \\

\hline
\end{tabular}
\end{center}
\caption{Flavour composition (in \%), in the Y-event analysis, of the 
anti-tagged, tagged and
  normal-mixture jet samples determined from the \jt\ Monte Carlo
  including simulation of the detector. The quark jet tagging used the
  parameters  $\theta = 7\degs$ and $\fcut = 0.75$. The errors are
  statistical only.}
\label{tag_flavour}
\end{table}

\begin{table}[htb]
\begin{center}
\begin{tabular}{|l|c|c|c|c|c|}
\hline

\multicolumn {1}{|c|}{Y-events} & \multicolumn {3}{|c|}{Absolute rates}& \multicolumn{2}{|c|}{Relative rates} \\

 \hline
                         & \rgqch\     & \rgqk\ & \rgql\  & \kzero\  & \lam  \\
\hline
Detector effects         & 0.003      & $<0.01$& 0.04  & $<0.01$ &  0.04\\
Quark jet  identification& 0.002      & 0.05   & 0.07  & $0.04$ & 0.06 \\ 
Event selection          & 0.010      & 0.03   & 0.08  & 0.02 & 0.08 \\ 
Jet purity determination & 0.003      & $<0.01$& 0.03 & $<0.01$ & 0.01 \\
Background determination & $-$        & 0.04   & 0.05 & 0.04 & 0.05\\ 
Efficiency determination & $<0.001$   & 0.04   & 0.11 & 0.04 & 0.10 \\
Monte Carlo statistics   & 0.005      & 0.03   & 0.07 & 0.03 & 0.07\\
\hline
Total systematic error   & 0.012      & 0.09   & 0.18  & 0.07 & 0.17\\
\hline
Statistical error        &  0.006     & 0.08   & 0.11 & 0.07 & 0.10 \\
\hline 
\end{tabular}
\end{center}
\caption{Breakdown of the contributions to the uncertainties on the
  ratios of particle production in quark and gluon jets from the 
Y-event analysis.}
\label{T_syst}
\end{table}

\begin{table}[htb]
\begin{center}
\begin{tabular}{|cc|c|c|c|}
  \hline
\multicolumn{2}{|c|}{Ratios of relative rates } &  OPAL Data & \jt~7.4 & \hw~5.9 \\
\hline
Energy-based & \kzero\ & $ 1.10 \pm 0.02 \pm 0.02 $ & 0.94 & 0.73 \\
(\ycut)      & \lam\   & $ 1.41 \pm 0.04 \pm 0.04 $ & 1.26 & 0.88 \\
\hline
Y-events     & \kzero\ & $ 0.94 \pm 0.07 \pm 0.07 $ & 0.95 & 0.62 \\
             & \lam\   & $ 1.18 \pm 0.10 \pm 0.17 $ & 1.34 & 0.87 \\
\hline
\hline
\multicolumn{2}{|c|}{Ratios of absolute rates } & OPAL Data & \jt~7.4 & \hw~5.9 \\
\hline
Y-events     & \kzero\ & $ 1.05 \pm 0.08 \pm 0.08 $ & 1.06 & 0.70 \\
             & \lam\   & $ 1.32 \pm 0.11 \pm 0.18 $ & 1.56 & 0.99 \\
             & charged & $ 1.116 \pm 0.006 \pm 0.012$& 1.16 & 1.13\\
\hline
\end{tabular}
\end{center}
\caption{Ratios of the relative \ks\ and \lam\ 
  production rates in gluon and quark jets from both analyses,
  together with the predictions of \jt~7.4 (the predictions of \jt~7.3
  correspond to those of \jt~7.4 for \kzero\ and are about 0.04
  lower for \lam) and \hw~5.9. The ratios of absolute rates determined in
  the Y-events analysis are also shown.}
\label{T_summary}
\end{table}

%
%
\clearpage

\end{document}